\documentclass{elsart}

\usepackage{graphics}
\usepackage{graphicx}
\usepackage{amssymb}

\newcommand {\CePrB}[2] {Ce$_{#1}$Pr$_{#2}$B$_6$}
\newcommand {\CeLaB}[2] {Ce$_{#1}$La$_{#2}$B$_6$}
\newcommand {\CeNdB}[2] {Ce$_{#1}$Nd$_{#2}$B$_6$}

\newcommand {\TQ} {\ensuremath{T_{\mathrm{Q}}}}
\newcommand {\TIC}[1] {\ensuremath{T_{\mathrm{IC}#1}}}
\newcommand {\V}[1] {\ensuremath{\mathbf{#1}}}
\begin{document}

\begin{frontmatter}

Version \today{}



\title{Magnetic phase diagram of \CePrB{0.70}{0.30}}
%

\author[LLB]{J.-M. Mignot\corauthref{JMM}}
\author[ADSM]{M. Sera},\,
\author[ADSM]{F. Iga}

\address[LLB]{Laboratoire L\'{e}on Brillouin, CEA-CNRS, CEA/Saclay, F-91191 Gif sur Yvette, France}  
\address[ADSM]{ADSM, Hiroshima University, Higashi Hiroshima, 739-8530, Japan}

\corauth[JMM]{Corresponding author. Tel: +33 1 6908 8708; fax: +33 1 6908 8261; \textit{Email address:} jmignot@cea.fr}

\begin{abstract}
Low-temperature, high-field ($H_{[1\bar{1}0]} \le 7.5$~T), neutron diffraction experiments on single-crystal Ce$_{0.70}$Pr$_{0.30}$$^{11}$B$_6$ are reported. Two successive incommensurate phases are found to exist in zero field. The appearance, for $H \ge 4.6$~T at $T = 2$~K, of an antiferromagnetic structure, $\V{k}_{\mathrm{AF}}=(\frac{1}{2},\frac{1}{2},\frac{1}{2})$, most likely due to an underlying antiferroquadrupolar order, is discussed in connection with recent x-ray diffraction experiments.
\end{abstract}

\end{frontmatter}


Light rare-earth (RE) hexaborides have been extensively studied because they realize a delicate balance between magnetic exchange and different types of higher-order intersite multipole interactions. Chemical substitution experiments make it possible to explore the effects of changes in relative interaction strengths, anisotropies due to local crystal-field (CF) potentials and/or intersite multipole couplings, and degeneracies of electronic ground states. Previous studies of \CeLaB{x}{1-x}\ \cite {Hiroi97} and \CeNdB{x}{1-x}\ \cite{Kobayashi2003} solid solutions have revealed pronounced differences in the pattern of ordered phases in the ($H$, $T$) plane, which can be traced back to one or more of the latter mechanisms. The \CePrB{x}{1-x} series is specific in the sense that Pr$^{3+}$ is a non-Kramers ion with a $\Gamma_5$ triplet CF ground state carrying no octupole moment.
It has been argued \cite{Kishimoto05} that its magnetic properties reflect a competition between $O_{xy}$-type antiferroquadrupolar (AFQ) and antiferromagnetic (AFM) exchange interactions, and denote a much stronger Ce-$RE$ coupling as compared to the Nd case. The concentration $x = 0.7$ studied in Ref.~\cite{Kishimoto05} is characterized by a strong suppression of the AFQ ordering temperature, together with an increase of the N\'eel temperature, as compared to pure CeB$_6$, resulting in the qualitatively new situation  $\TQ \ll \TIC{2} = 4.1$~K (where \TIC{2} denotes the higher magnetic phase transition temperature). In an external magnetic field applied parallel to a twofold cubic axis $\langle110\rangle$, a rich phase diagram is observed, with no less than 7 different phases  delineated from a combination of magnetization and magnetotransport experiments. 

In the present high-field neutron diffraction study, we have characterized the ordered magnetic structures which exist in different regions of the ($H$, $T$) plane. The phases are labeled hereafter according to the notations of Ref.~\cite{Kishimoto05}. Experiments were performed on the two-axis lifting detector diffractometer 6T2 (LLB, Saclay) using a PG-filtered incident neutron beam with wavelength $\lambda = 2.354$~\AA. The sample was a high-quality single crystal prepared from low-absorption $^{11}$B isotope.

\begin{figure}
	\begin{center}
	\includegraphics[width=0.6\columnwidth, angle=-90]{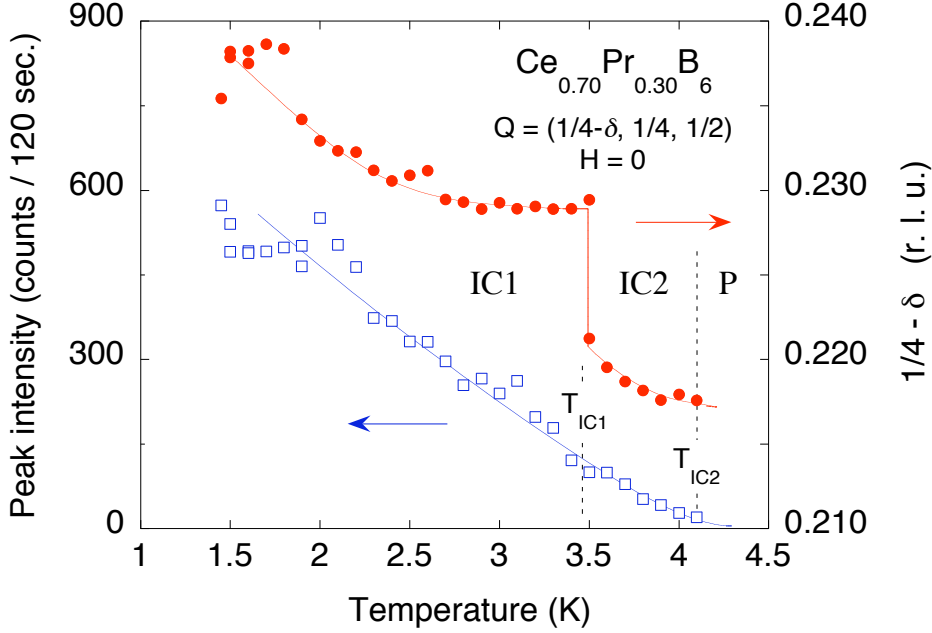}
	\end{center}
	\caption{Temperature dependence of the maximum intensity (left) and position
	(right) of the IC peak $(\frac{1}{4}-\delta)\frac{1}{4}\frac{1}{2}$ at $H = 0$.}
\label{fig1}
\end{figure}

In zero field, the first magnetic phase forming from the disordered state, IC2, is characterized by satellites corresponding to the incommensurate (IC) magnetic wavevector $\V{k}_{\mathrm{IC2}} = (\frac{1}{4}-\delta, \frac{1}{4},\frac{1}{2})$ with $\delta = 0.026 \pm 0.001$. With lowering temperature, $\delta$ decreases slightly down to \TIC{1} where it exhibits a discontinuous jump associated with the first-order transition into phase IC1 (Fig. \ref{fig1}). Quite remarkably, the wavevector in the low-temperature phase is still distinctly incommensurate, with $\delta$ close to 0.020  just below $\TIC{1} = 3.46$~K, and slowly decreasing to reach 0.012 at the minimum measuring temperature of $T_{\mathrm{min}} \approx1.5$~K. The transition at \TIC{1} cannot thus be typified as ``lock-in'', even though the deviation from commensurability of the structure in phase IC1 at $T_{\mathrm{min}}$ corresponds to periods of the order of 50 unit cells. The same type of IC wavevector was previously found in the IC phase of pure PrB$_6$, but the incommensurability in that case was much more pronounced, $\delta = 0.05$ \cite{Effantin85}.

\begin{figure}
	\begin{center}
	\includegraphics[width=0.6\columnwidth, angle=-90]{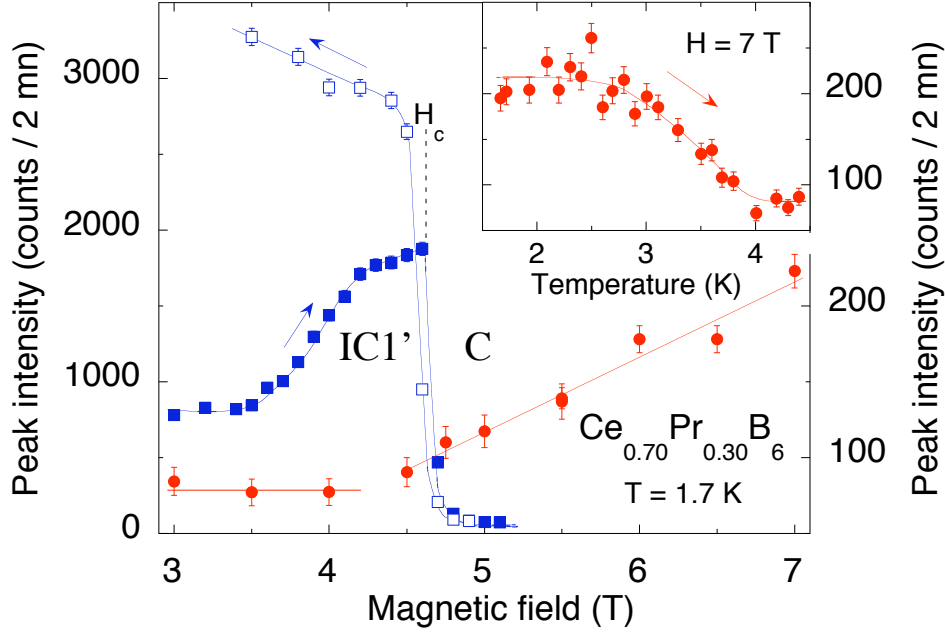}
	\end{center}
	\caption{Peak intensities of the $\frac{1}{2}\frac{1}{4}(\frac{1}{4}-\delta)$ (squares) and $\frac{1}{2}\frac{1}{2}\frac{1}{2}$ (circles) reflections as a function of field; inset: $T$ dependence of the latter intensity at $H = 7$~T.}
	\label{fig2}
\end{figure}

Ramping up the magnetic field $H \parallel [1\bar{1}0]$ produces $i$) no significant change in magnetic intensities and peak positions in phase IC1 up to $\approx 1$~T; $ii$) a pronounced redistribution of intensities among magnetic reflections associated with different branches of the star of $\V{k}_{\mathrm{IC1}}$, taking place within the so-called IC1' phase: specifically, the peaks $(\frac{1}{4}-\delta)\frac{1}{4}\frac{1}{2}$ and $\frac{1}{4}(\frac{1}{4}-\delta)\frac{1}{2}$, both corresponding to an AF stacking parallel to the fourfold cubic axis $[001]$ normal to the field, are fully suppressed above 1.8~T, whereas all others gain intensity; $iii$) a sudden drop in all incommensurate peak intensities at a critical field $H_{\mathrm{c}} = 4.62 \pm 0.05$~T, denoting the transition into phase C (Fig.~\ref{fig2}); correspondingly, a magnetic reflection appears, superimposed on a weak second-order nuclear peak, at the AFM position $\V{Q}_{AFM} = (\frac{1}{2}, \frac{1}{2},\frac{1}{2})$. It should be noted that no sizable change in the value of $\delta$ could be detected between $H = 0$ and $H_{\mathrm{c}}$. The domain repopulation effects observed in phase IC1' are reminiscent of those observed, for the same field direction, in the IC phase of PrB$_6$, and thus consistent with the same type of planar 2\V{k} structure as proposed in Ref.~\cite{Effantin85}. They also exhibit considerable hysteresis when the field is swept back to $H = 0$. The AFM component further increases with increasing $H$ from 5 to 7~T.

Heating up the sample in a fixed field of 7~T results in a gradual reduction of the AFM peak  intensity, starting around 2.5~K and extending all the way to $T = 4$~K, i. e. till the paramagnetic state is recovered (inset in Fig.~\ref{fig2}). It was also found by sweeping the field at $T = 3$~K that the intensity of the IC peak $\frac{1}{2}\frac{1}{4}(\frac{1}{4}-\delta)$ drops to zero upon entering the IC2' phase above 4.5~T. These results imply that, in the high-field regime, the staggered magnetic component revealed by the neutron data is dominated by the underlying AFQ order, which imposes its wavevector $\V{k}_{\mathrm{Q}} = (\frac{1}{2}, \frac{1}{2},\frac{1}{2})$ very much in the same way as in phase II of CeB$_6$. It is surprising, however, that no change could be detected in the diffraction pattern on crossing the transition lines between phases II, IC2', and C. The distinctive characteristics of these phases thus remain to be determined. If phase IC2' somehow represents an extension of phase IC2 into the AFQ ordered state, then it must be concluded that the incommensurability is destabilized by the AFQ order. Conversely, we have seen that, upon increasing $H$, the IC magnetic order existing in phase IC1 remains essentially unaffected (apart from domain population effects which may not be directly relevant here) when crossing the extension from high fields of the \TQ\ line, which marks the limit between phases IC1 and IC1'. This result is quite remarkable since recent nonresonant x-ray diffraction results on the same compound, published in these Proceedings \cite{Tanaka06}, clearly establish that AFQ order with the wavevector $\V{k}_{\mathrm{Q}}$ does exist in phase IC1', whereas the associated intensity drops drastically in phase C. Careful comparison of the neutron and x-ray diffraction results is essential to gain insight into the origin of this intriguing behavior.
\\
\\
We wish to thank J.-L. Meuriot for technical assistance in setting up  the cryomagnet.

%
%
%
%

\end{document}